\documentclass[a4paper,USenglish]{lipics-v2021}

\pdfoutput=1 
\hideLIPIcs  

\title{Memory-Anonymous Starvation-Free Mutual Exclusion:
Possibility and Impossibility Results}

\titlerunning{Memory-Anonymous Starvation-Free Mutual Exclusion} 

\author{Gadi Taubenfeld}{Reichman University, Herzliya, Israel \and \url{https://faculty.runi.ac.il/gadi}}{ }{https://orcid.org/0000-0003-3070-5370}{}

\authorrunning{G. Taubenfeld}

\Copyright{Gadi Taubenfeld}



\nolinenumbers 

\EventEditors{John Q. Open and Joan R. Access}
\EventNoEds{2}
\EventLongTitle{42nd Conference on Very Important Topics (CVIT 2016)}
\EventShortTitle{CVIT 2016}
\EventAcronym{CVIT}
\EventYear{2016}
\EventDate{December 24--27, 2016}
\EventLocation{Little Whinging, United Kingdom}
\EventLogo{}
\SeriesVolume{42}
\ArticleNo{33}

\begin{document}

\maketitle

\newcommand{\Xomit}[1]{ }
\newcommand{\abs}[1]{ }
\newcommand{\stretchvspace}{\vspace{0.5\abovedisplayskip}}
\newcommand{\notion}[2]{
\stretchvspace
\noindent
{\bf #1\/} {\em #2}
\stretchvspace
}




\newcommand{\pp}{p.i}
\def\false{{\it false}}
\def\true{{\it true}}
\def\myview{{\it myview}}
\def\mycounter{{\it mycounter}}
\def\mypointer{{\it mypointer}}
\def\mygo{{\it mygo}}
\def\waiting{{\it waiting}}



\begin{abstract}
In an anonymous shared memory system,
all inter-process communications are via shared objects; however, unlike in standard systems,
there is no a priori agreement between processes on the names of shared objects \cite{Tau2017podc,Tau2022JACM}.
Furthermore, the algorithms are required to be symmetric; that is,
the processes should execute precisely the same code, and the only way
to distinguish processes is by comparing identifiers for equality.
For such a system, read/write registers are called anonymous registers.
It is known that symmetric deadlock-free mutual exclusion is solvable
for any finite number of processes using anonymous registers \cite{AIRTW2019}.
The main question left open in \cite{Tau2017podc,Tau2022JACM} is the existence of
starvation-free mutual exclusion algorithms for two or more processes.
We resolve this open question for memoryless algorithms, in which a process that
tries to enter its critical section does not use any information about its
previous attempts.
Almost all known mutual exclusion algorithms are memoryless.
We show that,
\begin{enumerate}
  \item
There is a symmetric memoryless starvation-free mutual exclusion algorithm
for two processes using $m\geq 7$ anonymous registers if and only if $m$ is odd.
  \item
There is no symmetric memoryless starvation-free mutual exclusion algorithm for $n\geq 3$ processes
using (any number of) anonymous registers.
\end{enumerate}
Our impossibility result is the only example of a system with fault-free processes,
where global progress (i.e., deadlock-freedom) can be ensured, while individual progress to each process
(i.e., starvation-freedom) cannot.
It complements a known result for systems with failure-prone processes,
that there are objects with lock-free implementations but without wait-free implementations \cite{ACHP2022,HerSPAA91}.

%
\end{abstract}


\section{Introduction}
\subsection{Anonymous shared memory}

A central issue in distributed systems is coordinating
the actions of asynchronous processes. In the context where
processes communicate via reading and writing from shared memory,
in almost all published concurrent algorithms, it is assumed that
the shared memory locations have global names, which are
a priori known to all the participating processes.
The intriguing question of what and how coordination can be achieved without relying on
such lower-level agreement about the names of the memory locations was introduced
and studied in \cite{Tau2017podc,Tau2022JACM}.

We assume that all inter-process communications
are via shared read/write registers which are initially in a known state.
However, unlike in the standard model, from the point of view of the processes,
the registers do not have global names.
Such registers are called \emph{anonymous} registers.
Algorithms correct for a model where the registers are anonymous
are called \emph{memory-anonymous} algorithms.

There are fundamental differences between
the standard shared memory model
and the strictly weaker anonymous shared memory model \cite{Tau2022JACM}.
Besides enabling us to understand better the intrinsic
limits for coordinating the actions of asynchronous processes,
the anonymous shared memory model with symmetric processes
has been shown to be useful in modeling
biologically inspired distributed computing methods,
especially those based on ideas from molecular biology \cite{RTB2021}.
%
%
%
The main question left open in \cite{Tau2017podc,Tau2022JACM} is regarding
the existence of symmetric starvation-free mutual exclusion algorithms for two or more processes
when using anonymous registers.
In this article, we resolve this open question for memoryless algorithms.

\subsection{Mutual exclusion, symmetric algorithms, memoryless algorithms}
%
\textbf{The mutual exclusion problem.}
The mutual exclusion problem is to design an algorithm that guarantees
mutually exclusive access to a critical section among several competing
processes \cite{Dij65}.
It is assumed that each process executes a sequence
of instructions in an infinite loop. The instructions are divided into four
continuous sections: the remainder, entry, critical, and exit.
The exit section is required to be wait-free --
its execution must always terminate.
It is assumed that processes do not fail and
that a process always leaves its critical section.
The \textit{mutual exclusion problem} is writing the code for the entry and exit sections
to satisfy the following two basic requirements.
\begin{itemize}
\item
\emph{Deadlock-freedom:}
If a process tries to enter its
critical section, then some process, not necessarily the same one,
eventually enters its critical section.
\item
\emph{Mutual exclusion:}
No two processes are in their critical sections simultaneously.
\end{itemize}
The satisfaction of the above two properties is the minimum required for
a mutual exclusion algorithm.
For an algorithm to be fair, the satisfaction of the following stronger progress
condition is required.
\begin{itemize}
\item
\emph{Starvation-freedom:}
If a process is trying to enter its critical section, then this process
eventually enters its critical section.
\end{itemize}

\noindent
\textbf{Symmetric algorithms.}
A symmetric algorithm is an algorithm in which
the processes execute exactly the same code, and the only way to distinguish
processes is by comparing identifiers for equality.
A process can determine if two identifiers are the same,
but nothing else can be determined when they are different.
Identifiers can be written, read, and compared, but there is no way of
looking inside any identifier. Thus, for example, knowing whether
an identifier is odd or even is impossible.

Furthermore, (1) a process can only compare its identifier with another
and cannot compare it with a constant value, and (2)
the local variables of the different processes have the same names, and
(local variables with the same names) are initialized to the same values.
Otherwise, it would be possible to distinguish different processes.
In symmetric algorithms, as defined above, we say that the processes are symmetric.

As symmetric algorithms do not depend on an order relation between process identities,
they require fewer assumptions and are consequently
more general than non-symmetric algorithms.
%
The symmetry constraint on process identities can be seen as
the ``last step'' before process anonymity \cite{RT2022,RT2022b}.
Symmetric algorithms with non-anonymous memory have been investigated for years \cite{SP89}.
Following \cite{Tau2017podc}, we consider a model in which the memory is anonymous, and
the processes are symmetric.\\
\\
\textbf{Memoryless algorithms.}
A memoryless mutual exclusion algorithm is an algorithm in which
when all the processes are in their remainder section, the values of
all the registers (local and shared) are the same as their initial values.
This means that a process that
tries to enter its critical section does not use any information about its
previous attempts (like the fact that
it has entered its critical section five times so far).
Put another way, in a memoryless algorithm, processes have only a
single remainder state and hence cannot retain any memory of
prior executions of the algorithm.

All known mutual exclusion algorithms which use anonymous registers
and almost all known mutual exclusion algorithms that use non-anonymous registers
are memoryless. For example, Lamport's Bakery algorithm is memoryless \cite{Lam74}.
Memoryless mutual exclusion algorithms are usually simpler to understand
than those that are not memoryless.
Furthermore, memoryless algorithms can better handle system-wide failures (i.e., all processes
crash simultaneously), as upon recovery, the system can be initialized to
the (single) initial state, not affecting processes that had not participated
in the algorithm during the failure.
When using non-anonymous registers,
$2n-1$ registers are necessary and sufficient for designing
a symmetric memoryless starvation-free mutual exclusion algorithm for $n$ processes \cite{SP89}.
\subsection{Contributions: individual progress vs.\ global progress}
It is known that symmetric memoryless deadlock-free mutual exclusion is solvable
for any finite number of processes using anonymous registers \cite{AIRTW2019}.
The main question left open in \cite{Tau2017podc,Tau2022JACM} is the existence of
symmetric starvation-free mutual exclusion algorithms for two or more processes.
We resolve this open question for memoryless algorithms by proving the following
possibility and impossibility results,
\begin{enumerate}
  \item
There is a symmetric memoryless starvation-free mutual exclusion algorithm
for two processes using $m\geq 7$ anonymous registers if and only if $m$ is odd.
  \item
There is no symmetric memoryless starvation-free mutual exclusion algorithm for $n\geq 3$ processes
using (any number of) anonymous registers.
\end{enumerate}
Our possibility result shows that (1) there is no separation between deadlock-freedom and starvation-freedom
for two processes, and
(2) there is a separation between deadlock-freedom and starvation-freedom
for three or more processes.
These results enable
us to understand better the intrinsic
limits for achieving fairness between asynchronous processes.

Interestingly, as a byproduct of the proof of our impossibility result, we get a general
time complexity lower bound for every
symmetric \emph{deadlock-free} mutual exclusion algorithm for $n\geq 2$ processes
using $m$ anonymous registers.
Namely, a process must incur $\lceil m/2 \rceil$ remote memory references (RMR) to
enter and exit its critical section once.
The RMR complexity of our starvation-free algorithm is $O(m)$.

Two main progress conditions have been studied for asynchronous
shared-memory systems with failure-prone processes.
The first, wait-freedom, ensures individual progress to each process, i.e.,
its operations complete as long as it takes an infinite number of steps \cite{Her91}.
The second, lock-freedom, requires only global progress; namely, if a process takes
an infinite number of steps, then some (possibly other) processes complete
their operations \cite{HW90}.
Wait-freedom corresponds to starvation-freedom in fault-free systems,
while lock-freedom corresponds to deadlock-freedom.

It was shown in \cite{HerSPAA91} that there is an object for which there is a
lock-free implementation for two processes using only non-anonymous atomic registers, and there
is no wait-free algorithm in the same setting.
A more general result was presented in \cite{ACHP2022}, showing that such a separation
exists (1) also for more than two processes and (2) when primitives stronger
than atomic registers are used.
Our result, which shows a separation between deadlock-freedom and starvation-freedom
for three or more processes, complements these results.
The results together show that in shared memory systems where either
failures are possible or process symmetry and memory anonymity are assumed,
it is not always possible to ensure individual progress in situations where global progress is possible.
Thus, achieving various levels of fairness depends on the underlying system assumptions.

\section{Preliminaries}
\label{sec:Model}
\textbf{Processes.}
Our model of computation consists of a fully asynchronous collection
of $n$ deterministic processes that communicate via $m$ anonymous registers.
Asynchrony means that there is no assumption on the relative speeds of the processes.
Each process has a unique identifier, which is a positive integer.
Since we want to make as few assumptions as possible,
it is \emph{not} assumed that the identifiers of the $n$ processes
are taken from the set $\{1,...,n\}$.
Thus, a process does not a priori know the identifiers of the other processes.
The processes do know the values of $n$ and $m$.
We assume that processes do not fail.
As always assumed when solving the mutual exclusion problem, participation is not required -- a process may
stay in its remainder section and never move to its entry section.\\
\\
\textbf{Memory.}
The shared memory consists of $m$ \emph{anonymous} shared registers.
For $m$ anonymous registers, $r_1,...,r_m$,
the adversary can fix, for each process $p$, a permutation
${\pi}_p: \{r_1,...,r_m\}\rightarrow \{r_1,...,r_m\}$
of the registers such that, for process $p$, the $j$'th anonymous register is ${\pi}_p (r_j)$.
In particular, when process $p$ accesses its $j$'th anonymous register, it accesses ${\pi}_p (r_j)$.
Algorithms designed for such a system must be correct regardless of the permutations
chosen by the adversary.
The permutation fixed for process $p$ is called the naming assignment of $p$.

All the anonymous registers are assumed to be initialized to the same value.
Otherwise, thanks to their different initial values, it would be possible to
distinguish different registers, and consequently, the registers would no longer be fully anonymous.

With an \emph{atomic} register, it is assumed that operations on the
register occur in some definite order. That is, each operation is an indivisible action.
In the sequel, by \emph{registers}, we mean anonymous atomic read/write registers.
A read/write register is a shared register that supports (atomic)
read and write operations.
The fact that anonymous registers do not have global names implies that
only multi-writer multi-reader anonymous registers are possible.
Such registers can both be written and read by all the processes.\\
%
\\
\textbf{Known results.}
In \cite{Tau2017podc}, it has been proven that
a necessary and sufficient condition for the design of a
symmetric deadlock-free mutual exclusion algorithm
for two processes using anonymous registers is that the number of registers is odd.
\begin{theorem}[\cite{Tau2017podc}]
\label{thm:mutex:twoProcesses:Taub2017}
There is a symmetric deadlock-free mutual exclusion algorithm
for two processes using $m\geq 2$ anonymous registers if and only if $m$ is odd.
\end{theorem}
We will use this result for two processes later in the paper.
The \emph{only if} part of the above theorem is a special case of the
following more general result from \cite{Tau2017podc} (Theorem 3.4):
There is a symmetric deadlock-free mutual exclusion algorithm
for $n\geq 2$ processes using $m\geq 2$ anonymous registers
only if for every positive integer $1<\ell\leq n$,
$m$ and $\ell$ are relatively prime.
An optimal symmetric deadlock-free mutual exclusion algorithm
using anonymous registers that matches the above general space bound
for $n\geq 2$ processes was presented in \cite{AIRTW2019}.

\section{A starvation-free mutual exclusion algorithm
for two processes}
We show that, for two processes, it is possible to design a
starvation-free mutual exclusion algorithm. In the next section, we prove
this is impossible for three or more processes.
\begin{theorem}
\label{thm:mutex:twoProcesses}
There is a symmetric memoryless starvation-free mutual exclusion algorithm
for two processes using $m\geq 7$ anonymous registers if and only if $m$ is odd.
\end{theorem}
The \emph{only if} direction follows from
Theorem~\ref{thm:mutex:twoProcesses:Taub2017}
(proven in \cite{Tau2017podc}),
where it has been proven that (when using anonymous registers)
any symmetric deadlock-free mutual exclusion algorithm
for two processes must use an odd number of anonymous registers.
To prove the \emph{if} direction, we present in Figure~\ref{fig:mutexFor2processes}
a symmetric memoryless starvation-free mutual exclusion algorithm
for two processes using $m$ anonymous registers, where
$m$ is an odd number greater than or equal to 7.
The question of whether
a symmetric memoryless starvation-free mutual exclusion algorithm exists
for two processes using 3 or 5 anonymous registers is open.

As the $m$ registers do not have global names, each process independently
numbers them. We use the notation $\pp[j]$ to
denote the $j^{th}$ register according to process $i$'s numbering,
for $1\leq j\leq m$.
Recall that a process's identifier is a positive integer.

\subsection{An informal description of the algorithm}
A shared register is \emph{free} when its value is 0.
Initially, all the registers are free.
A register is owned by process $i$, when its value is $i$.
There are two ways for a process to get permission to enter its CS  (i.e., Critical Section).
The first way is when a process owns $m-2$ registers.
Initially, a process tries to own $m-2$ registers by writing its identifier into free registers.
If the process succeeds, it may enter its CS,
and when done, it releases (i.e., sets to 0)
all the registers it owns.
By design, a process will never own more than $m-2$ registers.

When there is contention, each process first tries to own
as many registers as possible, but no more than $m-2$ registers.
Thus, each process will always succeed in owning at least one register
(and not at least two registers, as explained in the sequel).
After it attempts to own $m-2$ registers,
if a process notices that it owns less than  $\lceil m/2\rceil$ registers,
it becomes a loser. A loser acts as follows: if it owns more than two registers,
it releases all its owned registers except two.
Otherwise, when a loser owns one or two registers, it releases no registers.
Then, the loser writes ``waiting'' into the (one or two) registers it owns and waits.
Since the waiting process owns at most two registers, the other process, the winner,
keeps on trying to own more registers until it eventually succeeds
in owning $m-2$ registers and gets permission to enter its CS.

When a winning process exits its CS, the winner releases all the $m-2$ registers it owns,
which, as explained below, will prevent it from entering its CS again before
a waiting process gets a chance to enter its CS. This guarantees that starvation-freedom is satisfied.
To guarantee that one process will not enter its CS twice while the other process is waiting,
when a process starts its entry code, it repeatedly scans the registers until
none of them has the value ``waiting,'' and only then may it proceed.

The second way a process can enter its CS is by waiting first.
A waiting process owns one or two registers with the value ``waiting.''
The waiting process waits until all the registers it does not own are released
(i.e., have the value 0). Once this happens, the waiting process may immediately
enter its CS. That is, it need not own additional registers.
Upon exiting its CS, the (previously waiting) process releases the (one or two)
registers it owns.

There is one very delicate possible race condition that should be avoided.
Assume there is contention; process $i$ writes its identifier into
$m-2$ registers, while process $j$ writes its identifier into
two registers. At that point, just before process $j$ writes ``waiting'' into
the two registers it owns, process $i$ enters and exits its CS, releases the $m-2$ registers
it owned and then attempts to enter its CS again. Process $i$ reads all the registers,
finds out that no process is waiting, and is ready to try to own $m-2$ free registers.
Now, process $j$ continues to write ``waiting'' into the two registers it owns,
finds out that all the other $m-2$ registers are free, and enters its CS.
Process $i$ is now scheduled, owns the $m-2$ free registers and enters its CS,
violating the mutual exclusion requirement.

It is easy to resolve this race condition while satisfying only deadlock-freedom.
Resolving it while still satisfying starvation-freedom is more challenging.
Our solution is as follows: the first thing that process $i$ is doing (upon entering
its entry code) is to own one register and only then check whether $j$ is waiting.
This will guarantee that either,
\begin{enumerate}
\item
process $i$ notices that $j$ is waiting, in which case $i$ releases its owned register
and waits until no register has the value ``waiting,'' letting the waiting process enter its CS first; or
\item
process $i$ notices that no process is waiting, in which case, after $j$ writes ``waiting'' into
its owned registers; it will find out that not all the other registers are free, and
will wait for process $i$ to enter its CS first.
\end{enumerate}
This solution resolves the race condition.

There is one additional thing that needs to be explained.
Why do we reserve \emph{two} registers for the waiting process, not just one?
The answer is that by reserving two registers, when ``waiting'' is written into two registers by some process,
at \emph{least one} of the two will not be overwritten by the other process.
Consider the following scenario. Let $r_1,r_2$ and $r_3$ be free registers.
Process $j$ writes into $r_1$,$r_2$ and is ready to write into $r_3$.
Process $i$ writes into all the $m-3$ other registers and is ready to write into $r_3$.
Process $j$ writes into $r_3$, finds out it is a loser (because $m\geq 7$), releases $r_1$,
and writes ``waiting'' into $r_2$ and $r_3$.
Now, process $i$ is activated and writes into $r_3$, leaving only one register
with the value ``waiting.''

Finally, we explain why the algorithm does not work when $m=5$ (or when $m=3$).
Assume $m=5$. Let $r_1,r_2,r_3,r_4$ and $r_5$ be the five free registers.
Consider the following scenario.
Process $i$ writes into $r_1$ and $r_2$ and is ready to write into $r_3$.
Process $j$ writes into $r_5$ and $r_4$ and is ready to write into $r_3$.
Process $i$ writes into $r_3$, finds out that it owns $m-2$ registers, and enters its CS.
Then, Process $j$ writes into $r_3$, finds out that it owns $m-2$ registers, and enters its CS,
violating the mutual exclusion requirement.
We point out that, in contrast, there is a symmetric memoryless deadlock-free mutual exclusion algorithm
for two processes using $m\geq 3$ anonymous registers when $m$ is odd.

\begin{figure}[!ht]
\small
\hrule
\vspace{0.05in}
\textsc{Code of process $i$} \hfill // $i\neq 0$
\begin{tabbing}
\hspace{1.5em}\=\hspace{1.5em}\=\hspace{6em}\=\kill
\textbf{Constant:}\\
\> $m$: an odd integer $\geq 7$  \`// \texttt{$m\geq 7$ is the \# of shared registers}\\
\textbf{Shared variables:}\\
\> $\pp[1..m]$: array of $m$ anonymous registers, of type integer + the symbol ``waiting,'' init.\ all 0 \\
\`// \texttt{$\pp[j]$ is the $j^{th}$ register according to process $i$'s numbering}\\
\textbf{Local variables:}\\
\> $\myview[1..m]$: array of $m$ variables, initially all 0 \\
\> $\mycounter$,$j$,$k$: integer, initially 0\\
\> $\mygo$: boolean, initially $\false$
\end{tabbing}
\vspace{-0.2in}
\begin{tabbing}
\hspace{1.5em}\=\hspace{1.5em}\=\hspace{1.5em}\=\hspace{1.5em}\=\hspace{1.5em}\=\kill
%
\`//\texttt{give priority to a waiting process}\\
1 \> \textbf{repeat}
$\mycounter\leftarrow \mycounter +1$
\textbf{until} $\pp[\mycounter]= 0$ \`//\texttt{looking for a zero entry}\\
%
2 \> $\pp[\mycounter]\leftarrow i$\`//\texttt{own one register}\\
3 \> \textbf{for} $j=1$ \textbf{to} $m$ \textbf{do} $\myview[j]\leftarrow \pp[j]$ \textbf{end for} \`//\texttt{read the shared array}\\
4 \> \textbf{if} $\exists j\in\{1,...,m\}: \myview[j]= \waiting$ \textbf{then} \`//\texttt{is other process waiting?} \\
5 \>\> \textbf{if} $\pp[\mycounter]=i$ \textbf{then} $\pp[\mycounter]\leftarrow 0$ \textbf{end if} \`//\texttt{release owned register} \\
6 \>\> \textbf{repeat}\`//\texttt{the other process is waiting}\\
7 \>\>\> \textbf{for} $j=1$ \textbf{to} $m$ \textbf{do} $\myview[j]\leftarrow \pp[j]$ \textbf{end for} \`//\texttt{read the shared array}\\
8 \>\> \textbf{until} $\forall j\in\{1,...,m\}: \myview[j]\neq\waiting$ \`//\texttt{wait for CS to be released}\\
9 \> \textbf{end if}\\
\\
10 \> \textbf{repeat}\`//\texttt{try to own $m-2$ registers}\\
11 \>\> \textbf{for} $k=1$ \textbf{to} $m$ \textbf{do}\`//\texttt{access the $m$ registers} \\
12 \>\>\> \textbf{if} $\pp[j]=0$ \textbf{then} \`//\texttt{try to own one more}\\
13 \>\>\>\> \textbf{for} $j=1$ \textbf{to} $m$ \textbf{do} $\myview[j]\leftarrow \pp[j]$ \textbf{end for} \`//\texttt{read the shared array}\\
14 \>\>\>\> \textbf{if} $i$ appears in less than $m-2$ of the entries of $\myview[1..m]$ \textbf{then}\\
15 \>\>\>\> $\pp[j]\leftarrow i$ \textbf{end if} \textbf{end if} \`//\texttt{own one more}\\
16 \>\> \textbf{end for}\\
\`//\texttt{lose or win?}\\
17 \>\> \textbf{for} $j=1$ \textbf{to} $m$ \textbf{do} $\myview[j]\leftarrow \pp[j]$ \textbf{end for} \`//\texttt{read the shared array}\\
18 \>\> \textbf{if} $i$ appears in less than $\lceil m/2\rceil$ of the entries of $\myview[1..m]$ \textbf{then}
       \`//\texttt{lose}\\
19 \>\>\> $\mycounter\leftarrow 0$\\
20 \>\>\> \textbf{for} $j=1$ \textbf{to} $m$ \textbf{do} \textbf{if} $\pp[j]=i$ \textbf{then} \`//\texttt{release all owned registers} \\
21 \>\>\>\> \textbf{if} $\mycounter =2$ \textbf{then} $\pp[j]\leftarrow 0$\`//\texttt{except two of them}\\
22 \>\>\>\>\> \textbf{else} $\pp[j]\leftarrow \waiting$; $\mycounter\leftarrow \mycounter +1$ \textbf{end if}\`//\texttt{signal waiting}\\
23 \>\>\> \textbf{end for}\\
24 \>\>\> \textbf{repeat} \`//\texttt{wait for CS to be released}\\
25 \>\>\>\> \textbf{for} $j=1$ \textbf{to} $m$ \textbf{do} $\myview[j]\leftarrow \pp[j]$ \textbf{od} \`//\texttt{read the shared array}\\
26 \>\>\> \textbf{until} $\forall j\in\{1,...,m\}: \myview[j]\in\{0,waiting\}$\`//\texttt{no sign from other process} \\
27 \>\>\> $\mygo\leftarrow\true$\`//\texttt{may enter CS}\\
28 \>\> \textbf{end if} \\
29 \> \textbf{until} $i$ appears in $m-2$ of the entries of $\myview[1..m]$ or $\mygo = \true$\\
30 \> \emph{critical section}\\
 \`\texttt{//release all owned shared registers} \\
31 \> \textbf{if} $\mygo= \true$ \textbf{then}
            \textbf{for} $j=1$ \textbf{to} $m$ \textbf{do}
            \textbf{if} $\pp[j]= \waiting$ \textbf{then} $\pp[j]\leftarrow 0$ \textbf{end if} \textbf{end for}\\
32 \>\> \textbf{else}
       \textbf{for} $j=1$ \textbf{to} $m$ \textbf{do}
       \textbf{if} $\pp[j]= i$ \textbf{then} $\pp[j]\leftarrow 0$ \textbf{end if} \textbf{end for}\\
33 \> \textbf{end if} \\
34 \> \textbf{set all local variables to their initial values}
\end{tabbing}
\vspace{-0.2in}
\caption{\small{A symmetric memoryless starvation-free mutual exclusion algorithm for two processes.}
\label{fig:mutexFor2processes}}
\vspace{0.05in}
\hrule
\normalsize
\end{figure}

\subsection{Correctness Proof}

\begin{lemma}
The algorithm satisfies mutual exclusion.
\end{lemma}

\proof
We assume to the contrary that both processes enter their CS
simultaneously and show that this leads to a contradiction.
There are two ways for a process to enter its CS:
(1) by observing that it owns $m-2$ registers, and
(2) by writing ``waiting'' into the registers it owns and
then waiting until all  the other registers have the value 0.
So, four possible combinations exist for having two processes in their CS simultaneously.
We show that none of them may happen.
Let's call the two processes $p$ and $q$.
\begin{enumerate}
\item
\emph{Both processes observe that they own $m-2$ registers and enter their CS.}
Once $p$ observes that it owns $m-2$ registers and enters its CS,
at most one of these $m-2$ registers may later be overwritten by $q$.
Thus, while $p$ is in its CS, $q$ may own at most 3 registers.
Since $m\geq 7$, $q$ cannot observe that it owns $m-2$ registers
and hence will not enter its CS, while $p$ is in its CS -- a contradiction.
\item
\emph{Both processes write ``waiting'' into the registers they own and later enter their CS.}
This may happen only if $p$ and $q$ observe that each one of them owns less than half of the registers.
However, if $p$ observes that it owns less than half of the registers,
it must be the case that $q$ owns more than half of the registers and will not
write ``waiting'' into the registers it owns -- a contradiction.
\item
\emph{$p$ observes that it owns $m-2$ registers and enters its CS,
while $q$ writes ``waiting'' into the registers it owns and \emph{later} enters its CS while $p$ is still in its CS.}
Once $p$ enters its CS, all the registers it owns are not free.
So, $q$ will not be able to proceed since not all the registers
it does not own are free as required -- a contradiction.
\item
\emph{$p$ writes ``waiting'' into the registers it owns,
waits until all  the other registers have the value 0
and enters its CS.
$q$ observe that it owns $m-2$ registers and enters its CS,
while $p$ is still in its CS.}
Since, after writing ``waiting,'' $p$ must observe that all the registers it does not own
have the value 0 before it may enter its CS;
it must be the case that $q$ has written into one of the registers (line 2),
after $p$ has written ``waiting.'' Thus, $q$ will notice (line 4)
that $p$ is waiting and will not proceed to its CS -- a contradiction.
\hfill\qed
\end{enumerate}
%
\begin{lemma}
The algorithm satisfies starvation-freedom.
\end{lemma}
\proof
There are three loops where a process can get stuck.
\begin{itemize}
\item \textbf{Loop 1:}
The repeat loop at lines 6-8, where a process waits
until none of the registers is set to waiting.
\item \textbf{Loop 2:}
The inner repeat loop at lines 24-26, where a (waiting) process waits
until all the registers are set to 0 or waiting.
\item \textbf{Loop 3:}
The outer repeat loop at lines 10-29 where a process waits
until it either owns $m-2$ registers or its local register
$\mygo$ is set to $\true$.
\end{itemize}
We show that a process cannot get stuck (i.e., loop forever) in
any of these loops, which implies starvation-freedom.
Let's call the processes $p$ and $q$.

Assume $p$ is waiting in loop 1, and cannot proceed. This means that (1) $p$ is not owning any of the registers, and
(2) that $q$ has set at least one of its owned registers to ``waiting'' (line 22).
By the time $q$ reaches loop 3 (line 24), all its owned registers are set ``waiting,''
and all the other registers are free. Thus, $q$ will exit the loop, set $\mygo$ to
$\true$ and enter its CS. Later, in its exit code $q$ will release its owned registers
and return to its remainder region. From that point on (even if $q$ will try to enter its CS again),
as long as $p$ is waiting in loop 1, no register will be set to waiting.
This means that the condition in line $8$ is evaluated to $\true$ and thus
$p$ can proceed beyond loop 1.

Assume $p$ is waiting in loop 2, and cannot proceed. This means that (1) $p$ has set the
(one or two) registers it owns to ``waiting'' and is not trying to own more registers, and
(2) $q$ owns at least one register. There are two cases to consider:
\begin{itemize}
\item
$q$ notices (in line 4) that $p$ is waiting, in which case $q$ releases its owned register
and waits until no register has the value ``waiting,'' letting the waiting process $p$ proceed beyond loop 2,
setting $\mygo$ to $\true$ and proceed beyond loop 3.
\item
$q$ notices that no process is waiting, in which case, after $p$ writes ``waiting'' into
its owned registers and waits in loop 2, there is nothing that prevents $q$ from owning $m-2$ registers
and entering its CS. Later, in its exit code $q$ will release its owned registers
and return to its remainder region. If $q$ does not try to enter its CS again, then
$p$ can proceed beyond loop 2 and loop 3. If $q$ does try to enter its CS again, then $q$ might acquire (line 3)
one register before $p$ notices that the register is free; however,
$q$ will later notice (in line 4) that $p$ is waiting, in which case $q$ will release its owned register
and waits until no register has the value ``waiting,'' letting the $p$ proceed beyond loop 2 and loop 3.
\end{itemize}

Assume $p$ is looping in loop 3, never waits in loop 2,  and cannot proceed.
Since $m$ is odd,
this means that $p$ owns more than half of the registers, while $q$ holds
less than half of the registers. Thus, $q$ will eventually release the registers
it owns, except at most two of them. This will enable $p$ to acquire $m-2$ registers
and proceed beyond loop 3.
\qed\\
\\
\textbf{RMR complexity.}
An operation that a process performs on a memory location is considered a remote memory reference (RMR)
if the process cannot perform the operation locally on its cache 
and must transact over the multiprocessor’s interconnection network to complete the
operation. RMRs are undesirable because they take long to execute and increase the interconnection
traffic. Our algorithm achieves the ideal RMR complexity of $O(m)$ for cache coherent machines.
(Distributed Shared Memory machines are irrelevant for anonymous shared memory systems.)
This means that a process incurs $O(m)$ number of RMRs to satisfy a request (i.e, to enter and exit the critical section once).
It follows from Observation~\ref{observation:RMR} (Section \ref{sec:impossibility}) that this bound is tight.

\section{An impossibility result}
\label{sec:impossibility}

In the previous section, we have shown that, for two processes, it is possible to design
a symmetric memoryless starvation-free mutual exclusion algorithm. Next, we show
this is impossible for three or more processes.
%
%
%
%
\begin{theorem}
\label{SFmutex:memoryless:impossibility}
There is no symmetric memoryless starvation-free
mutual exclusion algorithm for $n\geq 3$ processes using (any number of) anonymous registers.
\end{theorem}

\noindent
%
The main argument is that, under the appropriate assignment of names to registers,
there is a way to run two processes when they are in their remainder section, requiring them
to write to all registers before one of them can enter its critical section.
This is essentially accomplished by renaming unwritten registers on the fly so that if the two processes are about to write to the same unwritten register for the first time, then they end up writing to two distinct, unwritten ones.
Thus, it is possible to hide all the write operations of a third process,
which will prevent it from ever entering its critical section.
\subsection{Basic definitions and observations}
\label{subsec:basic}
We first prove some basic (but general) observations regarding the mutual exclusion problem.
All the lemmas and definitions in Subsection~\ref{subsec:basic} refer to one arbitrary
deadlock-free mutual exclusion algorithm for $n$ processes using read/write registers.
Here, we do not need to assume that the algorithm is starvation-free, symmetric, or memoryless,
nor do we need to assume that the registers are anonymous.

We will use the following notions and notations.
An \emph{event} corresponds to an atomic step performed by a process.
An algorithm's (global) state is entirely described by the values
of the (local and shared) registers and the values of the program counters of
all the processes.
A \emph{run} is a sequence of alternating states and events (also called steps).
For the purpose of the impossibility proof, it is more
convenient to define a run as a sequence of events omitting all the states
except the initial state.
Since the events and the initial state uniquely determine the states in a run,
no information is lost by omitting the states.

Each event in a run is associated with a specific process
that is {\em involved} in the event.
We will use $x$, $y$, and $z$ to denote runs.
When $x$ is a prefix of $y$, we denote by $(y-x)$
the suffix of $y$ obtained by removing $x$ from $y$.
Also, we denote by $x ; seq$ the sequence obtained by extending $x$
with the sequence of events $seq$.
Processes are \emph{deterministic};
that is, for every two runs $x;e$ and $x;e^{\prime}$
if $e$ and $e'$ are events by the same process, then $e = {e^{\prime}}$.

We will often use statements like ``in run $x$ process $p$ is in its
remainder'', and implicitly assume that there is a function that for
any run and process, lets us know whether a process is in its
remainder, entry code, critical section, or exit code.
Also, saying that an extension $y$ of $x$ involves only
process $p$, means that all events in $(y-x)$ involve only process $p$.
Finally, by a run we always mean a finite run, by a register
we mean a shared register, and by the value of register $r$ in run $x$,
we always mean, the value of $r$ at the end of $x$.
Our first definition captures when two runs are indistinguishable to a
given process.\\
\\
\textbf{Definition.}
\emph{Run $x$ \textbf{looks like} run $y$ to process $p$, if
the subsequence of all events by $p$ in $x$ is the same as in $y$,
and the values of all the registers in $x$ are the same as in $y$.}\\
\\
The looks like relation is an equivalence relation.%
\footnote{
The term \emph{looks like},
adopted from \cite{TauBook2006},
is also called \emph{indistinguishable}
in the literature.}
The next step  by a given process always depends on the
process's previous steps and the registers' current values.
The previous steps uniquely determine whether the next step is
a read or a write. The current values of the registers
determine what value will be read in case of a read step.
If two runs look alike to
process $p$, then the next step by $p$ in both runs is the same.

\begin{lemma}
\label{le:ext}
Let $x$ be a run which looks like run $y$ to every process in a set $P$.
If $z$ is an extension of $x$ which involves only processes
in $P$, then $y;(z-x)$ is a run.
\end{lemma}

\proof
By a simple induction on $k$ -- the number of events in $(z-x)$.
The basis when $k=0$ holds trivially.
We assume the lemma holds for $k\geq 0$ and prove for $k+1$.
Assume that the number of events in $(z-x)$ is $k+1$.
For some event $e$, it is the case that $z = z';e$.
Since the number of events in $(z' -x)$ is $k$,
by the induction hypothesis $y' = y;(z' -x)$ is a run.
Let $p\in P$ be the process which is involved in $e$.
Then, from the construction,
the runs $z'$ and $y'$ look alike to $p$,
which implies that the next step by $p$ in both runs is the same.
Thus, since $z = z';e$ is a run, also $y';e = y;(z-x)$ is a run.
\qed

We next define the notion of a hidden process.\footnote{%
The notion of a hidden process was first defined in \cite{BL93}.}
Intuitively, a process is hidden in a given run, if all the
steps it has taken since the last time it has been in its
remainder, communicate no information to the other processes.
We say that a write event $e_1$ is {\em overwritten} by event $e_2$
in a given run $r$ if $e_2$ is a write event that happens after $e_1$
in $r$, and both $e_1$ and $e_2$ are writing events to the same register.\\
\\
\textbf{Definition.}
\emph{For process $p$ and run $z$, let $z'$ be the longest prefix of
$z$ such that $p$ is in its remainder in $z'$.
Process $p$ is \textbf{hidden} in run $z$ if each event which $p$
is involved in $(z-z')$ is either: a read event,
or a write event that is overwritten (in $z$)
before any other process has read the value written.}\\
\\
We notice that a process is not hidden if it is involved in a write
event that is not later overwritten, even if the write does not change
the current value of a register.
Also, if a process is in its remainder in $z$ then it is
hidden in $z$, and thus initially, all the processes are hidden.
A hidden process looks just like a process halted in its remainder,
and hence no process can wait until a hidden process takes a step.

\begin{lemma}
\label{cs-hidding}
If a process $p$ is in its critical section in run $z$
then $p$ is not hidden in $z$.
\end{lemma}

\proof
Assume to the contrary that process $p$ is hidden and
is in its critical section in run $z$.
Let $z'$ be the longest prefix of $z$
such that $p$ is in its remainder in $z'$.
Since $p$ is {\it hidden} in run $z$,
it is possible to remove from $z$ all the events
in which $p$ is involved in $(z-z')$ and get a new run $y$.
The run $y$ looks like $z$ to all processes
other than $p$, and $p$ is in its remainder in $y$.
By the deadlock-freedom property, there is an extension of $y$
that does not involve $p$ in which some process $q\neq p$ enters
its critical section.
Since $y$ looks like $z$ to all processes other than $p$,
by Lemma \ref{le:ext}, a similar extension exists starting from $z$.
That is, $q$ can enter its critical section in an extension
of $z$, while $p$ is still in its critical section.
However, this violates the mutual exclusion property.
\qed

\noindent
It follows from Lemma~\ref{cs-hidding}
that a process must write before it enters its critical section.

\subsection{Anonymity}
\label{subsec:anonymity}
The lemma in Subsection~\ref{subsec:anonymity} refers to one arbitrary
deadlock-free mutual exclusion algorithm for $n$ processes using \emph{anonymous} registers.
Here, we do not need to assume that the algorithm is starvation-free, symmetric, or memoryless.
We denote by $e_p$ a (read or write) event which involves process $p$.
When $x$ is a run and ${\pi}_p$ is a naming assignment of $p$, we denote by
\begin{itemize}
\item
$x[p, r_i \leftrightarrow r_j]$ the sequence obtained by replacing
every read event of register $r_i$ by process $p$ in $x$
with a read event of register $r_j$ by $p$ which returns the same value (as the event of reading $r_i$), and vice versa.
\item
${\pi}_p [r_i \leftrightarrow r_j]$ is the naming assignment where,
(1) ${\pi}_p [r_i \leftrightarrow r_j] (r_i) = {\pi}_p (r_j)$;
(2) ${\pi}_p [r_i \leftrightarrow r_j] (r_j) = {\pi}_p (r_i)$; and
(3) for every $k\notin\{i,j\}$, ${\pi}_p [r_i \leftrightarrow r_j] (r_k) = {\pi}_p (r_k)$.
\end{itemize}
Recall that all the anonymous registers are initialized to the same value.

\begin{lemma}
\label{lemma:anonymity}
Let $x$ be a run, $p$ a process, ${\pi}^{x}_p$ the naming assignment of $p$ (used in the run $x$),
and $r_i$ and $r_j$ two registers that were never written (by any process) in $x$.
Then, $y = x[p, r_i \leftrightarrow r_j]$ is a run, where the naming assignment
for $p$ (used in $y$) is ${\pi}^{x}_p [r_i \leftrightarrow r_j]$, and for the other processes the
naming assignments are the same as in $x$.
Furthermore, if $x;e_p$ is a run where $e_p$ is an event of writing the value $v$ into $r_i$
then $y;e^{\prime}_p$ where $e^{\prime}_p$ is an event of writing the value $v$ into $r_j$
is also a run.
\end{lemma}

\proof
Since $r_i$ and $r_j$ were never written in $x$, a read event by $p$
or any other process of each of those registers in $x$ would return the
initial value. So, swapping the read events of $r_i$ and $r_j$ by process $p$,
may only affect process $p$.

As for $p$, before its first event in $x$,
the adversary has fixed, for process $p$, a naming assignment
${\pi}^{x}_p: \{r_1,...,r_m\}\rightarrow \{r_1,...,r_m\}$
of the registers such that, for process $p$, the $j$'th anonymous register is ${\pi}^{x}_p (r_j)$.
In particular, when process $p$ accesses its $j$'th anonymous register, it accesses ${\pi}^{x}_p (r_j)$.
Let us define the permutation ${\pi}^{y}_p: \{r_1,...,r_m\}\rightarrow \{r_1,...,r_m\}$ as follows,
(1) ${\pi}^{y}_p (r_i) = {\pi}^{x}_p (r_j)$;
(2) ${\pi}^{y}_p (r_j) = {\pi}^{x}_p (r_i)$; and
(3) for every $k$ where $k\notin\{i,j\}$, ${\pi}^{y}_p (r_k) = {\pi}^{x}_p (r_k)$.
That is, whenever $p$ accessed $r_i$ before it will now access $r_j$ and vice versa.
Now, consider a run in which the processes are scheduled exactly in the order as
they are scheduled in $x$,
the naming assignment fixed (by the adversary) for $p$ is ${\pi}^{y}_p$,
and the naming assignments (in $y$) of the other processes are as in $x$.
By construction, the resulting run is run $y$.
Furthermore, if the next event by $p$ in
$x$ is of writing the value $v$ into $r_i$ then, since ${\pi}^{y}_p (r_i) = {\pi}^{x}_p (r_j)$,
the next event by $p$ in $y$ is of writing the value $v$ into $r_j$.
\qed

\subsection{The notions of a symmetric run and a symmetric state}
\label{subsec:SymmetricRun}
%
A (global) state is entirely described by the values
of the local and shared registers and the values of the program counters of
all the processes.

Intuitively, a state $\sigma$ is symmetric w.r.t. two processes if the ``subjective views''
of the processes at $\sigma$ are the same.
In the standard non-anonymous model, once two (symmetric) processes write
their ids (one after the other) into the same register, say $r_1$, their views after these two writes
are completed are no longer the same, if they inspect the current state,
one process will see that the value of $r_1$ equals its id,
while the other will see that the value is different from its id.
In the anonymous model, it is possible for each process
to see that the value of the first register according to its naming assignment
equals its id.

Below we define this notion more formally.
Let $m$ be the number of registers;
let $val_{\sigma} (r)$ be the value of register $r$ in state $\sigma$,
and assume that the names of the local variables of the processes are the same.
To distinguish the local variables
of the different (symmetric) processes, we will add the process id as a subscript to the variable names.\\
\\
\textbf{Definition.}
\emph{
Let $\sigma$ be a state
and let ${\pi}_p$ and ${\pi}_q$ be the naming assignments of
$p$ and $q$, respectively.
State $\sigma$ is \textbf{symmetric} w.r.t.\ $p$ and $q$ and their naming assignments ${\pi}_p$ and ${\pi}_q$,
if for every $1\leq k\leq m$
either,
\begin{itemize}
  \item
$val_\sigma({\pi}_p (r_k)) = val_\sigma({\pi}_q (r_k))$ and $val_\sigma({\pi}_p (r_k))\notin \{p,q\}$, or
  \item
$val_\sigma({\pi}_p (r_k)) = p$ and $val_\sigma({\pi}_q (r_k)) = q$, or
  \item
$val_\sigma({\pi}_p (r_k)) = q$ and $val_\sigma({\pi}_q (r_k)) = p$.
\end{itemize}
Furthermore, in $\sigma$, for every local variable, say local, either
(1) $local_p = local_q$ and the value of $local_p$ is not in $\{p,q\}$, or (2) $local_p = p$ and $local_q =q$, or (3) $local_p = q$ and $local_q =p$.
}\\
\\
\textbf{Definition.}
Run $x$ is \textbf{symmetric} w.r.t.\ $p$ and $q$ and their naming assignments ${\pi}_p$ and ${\pi}_q$,
if the state at the end of $x$ is symmetric
w.r.t.\ $p$ and $q$ and their naming assignments ${\pi}_p$ and ${\pi}_q$.\\
\\
The following lemma follows immediately from the definitions of a symmetric run and a symmetric state.
\begin{lemma}
\label{observation:basic:symmetric}
When processes $p$ and $q$ are symmetric (and deterministic),
if run $x$ is symmetric w.r.t.\ $p$ and $q$ and their naming assignments ${\pi}_p$ and ${\pi}_q$, then
\begin{enumerate}
  \item
either the next step of both processes is a read or the next step of both is a write;
  \item
if $p$ accesses ${\pi}_p (r_k)$ in its next step, then
$q$ accesses ${\pi}_q (r_k)$ in its next steps;
  \item
if $x$ is extended by a read event of $p$ followed by a read event of $q$, then
the resulting run is symmetric w.r.t.\ $p$ and $q$ and their naming assignments;
  \item
if $x$ is extended by a write event of $p$ followed by a write event of $q$ then
the resulting run is symmetric w.r.t.\ $p$ and $q$ and their naming assignments,
provided that $p$ and $q$ do not write into the same physical location;
and
  \item
$p$ and $q$ cannot be in their critical sections at (the end of) $x$.
\end{enumerate}
\end{lemma}

\begin{lemma}
\label{observation:sigma:symmetric}
In a symmetric deadlock-free mutual exclusion algorithm for $n\geq 2$ processes
using anonymous registers, the initial state
where all the processes are in their remainder sections is
symmetric w.r.t.\ every two processes and their naming assignments.
\end{lemma}

\proof
The proof follows immediately from the fact that in anonymous shared memory,
all the anonymous registers are initialized to the same value; and
that when the processes are symmetric,
the local variables of the different processes have the same names and
(local variables with the same names) are initialized to the same values.
\qed\\
\\
To simplify the presentation, for the rest of the section,
by a \emph{symmetric run} (resp.\ \emph{symmetric state}), we always mean
a symmetric run (resp.\ symmetric state) w.r.t.\ to every two processes and their naming assignments.
\subsection{Symmetry and anonymity}
\label{subsec:SymmetryAnonymity}
The following lemma in Subsection~\ref{subsec:SymmetryAnonymity} refers to one arbitrary
symmetric deadlock-free mutual exclusion algorithm for $n\geq 2$ processes
using $m$ \emph{anonymous} registers.
Here, there is no need to assume that the algorithm is starvation-free or memoryless.

A \emph{quiescent} state is one in which all the processes are in their remainder sections.
A memoryless algorithm is an algorithm that has exactly one quiescent state (which is the initial state).
When the possible number of quiescent states of a symmetric starvation-free mutual exclusion algorithm is more than one, by Lemma~\ref{observation:sigma:symmetric},
the initial (quiescent) state is  symmetric w.r.t.\ every two processes and their naming assignments.
However, for a non-memoryless algorithm,
except for the initial state, the other quiescent states are
not necessarily symmetric.

\begin{lemma}[main technical lemma]
\label{imp:lemma:main}
For every two processes $p$ and $q$, and every symmetric quiescent state $\sigma$,
there exist
naming assignments ${\pi}_p$ and ${\pi}_q$ for $p$ and $q$,
and a run $\rho$ with the following properties,
\begin{enumerate}
\item
$\rho$ starts from the state $\sigma$
and ends in some quiescent state,
\item
during $\rho$, only $p$ and $q$ take steps, and they enter and exit their critical sections once,
\item
during $\rho$, each one of the two processes writes into $\lceil m/2 \rceil$ different registers, and
\item
during $\rho$, each one of the $m$ anonymous registers is written at least once.
\end{enumerate}
\end{lemma}
%
\proof
It follows from
Theorem~\ref{thm:mutex:twoProcesses:Taub2017}
that (when using anonymous registers)
any symmetric deadlock-free mutual exclusion algorithm
for two processes must use an odd number of anonymous registers.
So, for the rest of this proof, we assume that $m$ is odd.

We prove the lemma by running the two processes $p$ and $q$,
starting from the (symmetric) state $\sigma$,
keeping the run symmetric (i.e., without breaking symmetry).
As long as symmetry is not broken,
by Lemma~\ref{observation:basic:symmetric},
none of the processes can enter its CS, because if it does, then the other process may enter
its CS as well, violating the mutual exclusion requirement.
Each time two more registers are written until only one is left.
At that point, the two processes try to write this last register.
Only when this last register is written is symmetry broken,
all the $m$ anonymous registers are written,
and we are done.

Initially, the adversary fixes for process $p$, the identity permutation
${\pi}_p (r_k) = r_k$ (where $1\leq k\leq m$), and fix for $q$ its reverse permutation ${\pi}_q (r_k) = r_{m-k+1}$.
For the sake of explanation, we arrange the registers in pairs, where the $k^{th}$ pair is (${\pi}_p (r_k)$, ${\pi}_q (r_k)$),
where $1\leq k \leq m$. In this pairing, each pair includes two different registers
except the $\lceil m/2 \rceil$ pair in which the register $r_{\lceil m/2 \rceil}$ appears twice.
Each other register appears in two pairs, and the other register is the same in those two pairs.
For example, when $m=5$, the pairs are ($r_1,r_5$), ($r_2,r_4$), ($r_3,r_3$),($r_4,r_2$), and($r_5,r_1$).

Next, we run the processes in \emph{lock-steps}. Each lock-step includes a step by $p$ followed by a step
of $q$. We observe that, as long as the constructed run is symmetric,
in one lock-step, if $p$ accesses $r_i$ and $q$ accesses $r_j$,
then (1) the pair ($r_i,r_j$) appears in the pairing described above, and (2)
by Lemma~\ref{observation:basic:symmetric},
both processes either read $r_i$ and $r_j$ or write into them.
Because of the initial symmetry, none of the processes may enter its CS without writing first.
(this also follows from Lemma~\ref{cs-hidding}).

The run $\rho$ is constructed by iteratively executing the following procedure until
all the $m$ registers are written at least once, as required.
Let ${\pi}_p$ and ${\pi}_q$ denote the current naming assignments of $p$ and $q$.
Let us denote by $x_{i-1}$, the run constructed so far, before the beginning of the $i^{th}$ iteration.
As we will see, by construction, $x_i$ will be a symmetric run for all $i\geq 0$.
By assumption, the quiescent state $\sigma$ is symmetric, and thus,
the run $x_0$, where all the processes are still in their remainder sections, is symmetric.%
\footnote{We notice that in the special case where the algorithm is memoryless,
there is no need to assume that $\sigma$ is symmetric, as
by Lemma \ref{observation:sigma:symmetric}, the single initial (quiescent) state $\sigma$ is symmetric.
}
\begin{quote}
\textbf{Iteration $i\geq 1$ begins here.}
As explained, $x_{i-1}$ is the symmetric run constructed so far.
First, $p$ and $q$ run (in lock-steps) until they are ready to write into $r_i$ and $r_j$, respectively.
Recall that by Lemma~\ref{observation:basic:symmetric}, either
both steps are read or both are write, in each lock-step.
Thus, to break symmetry they must eventually try to write.
Let's denote by $y_{i-1}$ the run just before the two writes.
Clearly, if $x_{i-1}$ is symmetric, then so is $y_{i-1}$.
There are four possible cases,
\begin{enumerate}
\item
$r_i$ and $r_j$ are different registers.
In this simple case, we let both processes complete their write operations.
The run $x_i$ is the run just after these two writes.
If $x_{i-1}$ is symmetric, then so is $x_i$.
The $i^{th}$ iteration completes here, and we are ready to start the next $i+1$ iteration.
\item
$r_i$ and $r_j$ refer to the same register, which has been written before.
The construction ensures this situation never happens (see Case 4 below).
\item
$r_i$ and $r_j$ refer to the same register which has not been written before,
and all other $m-1$ registers have already been written.
In this case, we let both processes complete their write operations. At this point,
all the $m$ registers have been written as requested. So we let the processes continue
running until, as guaranteed by the deadlock-freedom property, each one of them will eventually
enter its CS (but not simultaneously), return to its remainder section, and we are done.
The constructed run is $\rho$, and
\textbf{the construction of the run $\rho$ terminates here.}
\item
$r_i$ and $r_j$ refer to the same register which has not been written before,
and \emph{not} all other $m-1$ registers have already been written.
This is the more challenging case.
For some $k$, let ${\pi}_p (r_k)$ and ${\pi}_q (r_k)$ be two registers (different from $r_i$)
that have not been written so far (we do not care whether ${\pi}_p (r_k)$ and ${\pi}_q (r_k)$ have or have not
been read so far).
By Lemma~\ref{lemma:anonymity}, $z^1_{i-1} = y_{i-1}[p, r_i \leftrightarrow {\pi}_p (r_k)]$ is a legal run,
assuming ${\pi}_p$ is replaced with ${\pi}_p[r_i \leftrightarrow {\pi}_p (r_k)]$.
Furthermore, by applying lemma~\ref{lemma:anonymity} again,
$z^2_{i-1} = z^1_{i-1}[q, r_j \leftrightarrow {\pi}_q (r_k)]$ is a legal run,
assuming ${\pi}_q$ is replaced with ${\pi}_q[r_i \leftrightarrow {\pi}_q (r_k)]$.
Since $r_i$, ${\pi}_p (r_k)$ and ${\pi}_q (r_k)$ have never been written yet,
if $x_{i-1}$ is symmetric, so is $z^2_{i-1}$.

Furthermore, at (the end of) $z_2$,
$p$ and $q$ are ready to write into the two registers ${\pi}_p (r_k)$ and ${\pi}_q (r_k)$, respectively.
We now continue with run $z^2_{i-1}$, and
let both processes complete their write operations into ${\pi}_p (r_k)$ and ${\pi}_q (r_k)$.
Again symmetry is preserved.
The resulting run, after the writes is $x_i$ with which we continue to the next iteration.
It is important to notice that in this case (unlike in the other two cases)
$x_i$ is \emph{not} an extension of $x_{i-1}$, it is a completely new run.
In addition, in the next round, which starts with $x_i$, we use the (updated)
naming assignments
${\pi}_p[r_i \leftrightarrow {\pi}_p (r_k)]$ and
${\pi}_q[r_j \leftrightarrow {\pi}_q (r_k)]$ for processes $p$ and $q$.
As explained $x_i$ is symmetric w.r.t.\ $p$ and $q$ and their updated
naming assignments.
\end{enumerate}
\end{quote}
In case 4, we switch to a new run and new naming assignments.
We emphasize that, during a specific run, we do not change the naming assignments associated with
that run. We change them only when we switch to a completely new run.
The run $\rho$ that we end up with at the end of the construction uses fixed
naming assignments for $p$ and $q$ from the beginning to the end.
This whole construction can be viewed as searching for a run $\rho$ and fixed
naming assignments ${\pi}_p$ and ${\pi}_q$ for which \emph{Case 4} will never happen,
and this is exactly what we end up with. That is, once run $\rho$ is constructed with its
associated naming assignments for $p$ and $q$, if we consider
execution of run $\rho$ starting from $\sigma$ with the associated naming assignments,
then, by construction, the only time $p$ and $q$ will try to write
into the same register, would be \emph{after}
all other $m-1$ registers have already been written.
\qed

\begin{observation}
\label{observation:RMR}
The RMR complexity of every
symmetric deadlock-free mutual exclusion algorithm for $n\geq 2$ processes
using $m$ anonymous registers is at least $\lceil m/2 \rceil$.
\end{observation}

\proof
Let $\sigma$ be the initial state.
By Lemma \ref{observation:sigma:symmetric}, the initial state $\sigma$ is symmetric.
By Lemma~\ref{imp:lemma:main}, there exists a run
which starts from $\sigma$, in which exactly two processes participate
during which they enter and exit their critical sections once, and
each one of the two processes writes into $\lceil m/2 \rceil$ different registers.
The result follows.
\qed

\subsection{Proof of Theorem~\ref{SFmutex:memoryless:impossibility}}
\label{subsection:proof:mainTheorem}
%
Assume to the contrary that there is a symmetric memoryless \emph{starvation-free} mutual exclusion algorithm for $n\geq 3$ processes using anonymous registers. Let's call this algorithm $A$.
Let us denote by  $\sigma$ the single possible initial state.
By Lemma \ref{observation:sigma:symmetric}, $\sigma$ is symmetric.
Let $p_1, p_2$ and $q$ be three processes.
Using Lemma \ref{imp:lemma:main}, we will reach a contradiction by hiding all the write operations of $q$,
which, by Lemma~\ref{cs-hidding}, will prevent $q$ from ever entering its critical section.

This is done as follows. Assume all the processes are in their remainder sections;
thus, the current state is $\sigma$.
Now, process $q$ tries to enter its critical section. Before doing so, $q$ should execute its entry section
which must, by Lemma~\ref{cs-hidding}, involve at least one write operation.
So, we run $q$ alone until it is about to execute its first write operation.
Let's call the first register that $q$ is about to write register $r_1$.
Since $q$ has not modified any register yet, all the processes except $q$ cannot distinguish the current state
from the state $\sigma$ where all the processes (including $q$) are still in their remainder sections.
Thus, by Lemma~\ref{le:ext} and Lemma \ref{imp:lemma:main}, there is an extension of the current run which involves only
$p_1$ and $p_2$, in which $p_1$ and $p_2$ enter and then exit their critical sections once, and in that extension, each one of the anonymous registers is written (by either $p_1$ or $p_2$) at least once. We slightly modify this extension
of $p_1$ and $p_2$ by stopping them just before writing into register $r_1$; let $q$ complete a write operation into $r_1$; then let $p_1$ and $p_2$ overwrite the value written by $q$, and continue until $p_1$ and $p_2$
return to their remainder sections.

Since the write of $q$ into $r_1$ was immediately overwritten, $q$ is hidden. Hence
all the processes except $q$ cannot again distinguish the current state from the state
$\sigma$ where all the processes (including $q$) are still in their remainder sections.
Notice that here we use the assumption that the algorithm is memoryless,
and this will enable us to use Lemma~\ref{imp:lemma:main} repeatedly.
It is important to understand that since we have already used
Lemma~\ref{imp:lemma:main} once in the proof, the naming assignments
for processes $p_1$ and $p_2$ have been fixed, and these assignments can not be changed after that.
However, because of the memoryless assumption, this does not prevent us from applying
Lemma~\ref{imp:lemma:main} again to $\sigma$ using processes $p_1$ and $p_2$.

We notice that \emph{without} the memoryless assumption, after $p_1$ and $p_2$ return to their remainder sections,
it is no longer necessarily true that
all the processes except $q$ cannot again distinguish the current state from the state $\sigma$. This is so
because there might be another state $\sigma^{\prime}$ where all the processes are in their remainder sections.
Hence, in such a case (without the memoryless assumption), it would not be possible to apply Lemma \ref{imp:lemma:main} again.

By Lemma~\ref{cs-hidding}, at this point,
since $q$ is hidden,
$q$ must write again before it may enter its critical section.
So, we run $q$ alone until it is about to execute its second write operation.
Let's call the register that $q$ is about to write register $r_2$. (Register $r2$ might denote a different or the same register as $r_1$.)
As before, all the processes except $q$ cannot distinguish the current state from the state $\sigma$.
Thus, again by Lemma \ref{imp:lemma:main}, there is an extension of the current run that involves only  process
$p_1$ and $p_2$, in which $p_1$ and $p_2$ enter and then exit their critical sections once, and in that run each one of the anonymous registers is written at least once. We slightly modify this extension
of $p_1$ and $p_2$ by stopping them just before writing register $r_2$; let $q$ complete a write operation into $r_2$; then let $p_1$ and $p_2$ overwrite the value written by $q$, and continue until they return to their remainder sections. Since the write of $q$ into $r_2$ was also immediately overwritten, $q$ is still hidden, and thus
all the processes except $q$ cannot distinguish the current state from $\sigma$.

We can apply the above procedure as often as necessary, hiding all the write operations of $q$, which
by Lemma~\ref{cs-hidding},
prevents $q$ from ever entering its critical section. Thus, algorithm $A$ does not satisfy starvation-freedom
as promised. A contradiction. \qed

\subsection{A generalization}
For a non-memoryless algorithm,
except for the initial state, the other quiescent states are
not necessarily symmetric.
Under the assumption that all the quiescent states are symmetric,
it is possible to prove the following generalization of
Theorem~\ref{SFmutex:memoryless:impossibility}.
\begin{theorem}
\label{SFmutex:quiescent:impossibility}
There is no symmetric starvation-free mutual exclusion algorithm,
with at most $s$ quiescent states, for $n\geq 2s+1$ processes
using (any number of) anonymous registers,
assuming all the quiescent states are symmetric.
\end{theorem}

\proof
We slightly modify the proof of the impossibility result in Subsection~\ref{subsection:proof:mainTheorem}.
We assume to the contrary that there is a symmetric \emph{starvation-free} mutual exclusion algorithm
for $n\geq 2s+1$ processes using anonymous registers.
With each quiescent state $\sigma_i$ we associate two processes
$p^i_1$ and $p^i_2$. Let $q$ be a process not associated with any quiescent state.
Using the above-modified version of Lemma \ref{imp:lemma:main}, we can reach a contradiction
by hiding all the write operations of $q$, which, by Lemma~\ref{cs-hidding}, will prevent
$q$ from ever entering its critical section.

This is done as follows. Assume that all the processes are in their remainder sections,
and the current state is $\sigma_i$ which is assumed to be symmetric
(this is required for applying the modified version of Lemma~\ref{imp:lemma:main}).
Now, process $q$ tries to enter its critical section. Before doing so,
$q$ must execute at least one write operation.
So, we run $q$ alone until it is about to execute a write operation.
Let's call the register that $q$ is about to write, register $r$.
Since $q$ has not modified $r$ yet, all the processes except $q$ cannot distinguish the current state
from the quiescent state $\sigma_i$.
Thus, there is an extension of the current run, which involves only
$p^i_1$ and $p^i_2$, in which $p^i_1$ and $p^i_2$ enter and then exit their critical sections once, and in that extension, each one of the anonymous registers is written at least once. We slightly modify this extension
by stopping $p^i_1$ and $p^i_2$ just before writing into register $r$; let $q$ complete a write operation into $r$;
then let $p^i_1$ and $p^i_2$ overwrite the value written by $q$, and continue until $p^i_1$ and $p^i_2$
return to their remainder sections.

Since the write of $q$ into $r$ was immediately overwritten, $q$ is still hidden, and thus
all the processes except $q$ cannot distinguish the current state from one of the $s$ quiescent states.
We can apply the above procedure as often as necessary, hiding all the write operations of $q$, which
by Lemma~\ref{cs-hidding},
prevents $q$ from ever entering its critical section. Thus, the algorithm does not satisfy starvation-freedom
as promised -- a contradiction.
\qed

\section{Discussion}
We have shown that, while for two processes, it is possible to design a
symmetric memoryless starvation-free mutual exclusion algorithm using
anonymous registers, this is impossible for three or more processes.
These results imply that, while there is no separation between deadlock-freedom
and starvation-freedom for two processes, such a separation
between deadlock-freedom and starvation-freedom exists for three or more processes.
Thus, in anonymous shared memory systems where process symmetry is assumed,
it is impossible always to ensure individual progress in situations where global progress is possible.
This is the first known case of fault-free systems,
demonstrating a separation between individual and global progress.



%

\end{document}